\documentclass[final,authoryear,12pt]{elsarticle}

\usepackage{graphicx}
\usepackage{epstopdf}
\usepackage{hyperref}
\usepackage{setspace}
\usepackage{booktabs}
\usepackage{natbib}
\usepackage{bbm, dsfont}
\usepackage{amssymb}
 \usepackage{amsthm}
 \usepackage{amsmath}
\usepackage{amsfonts}
 \usepackage{geometry}
 \usepackage{bm}
\usepackage{color}
\usepackage{hyperref}
\usepackage{array,multirow}
\usepackage{caption}

\usepackage{lscape}

\usepackage{color,soul}
\definecolor{lightblue}{rgb}{.80,.95,1}
\sethlcolor{lightblue}
\setulcolor{red}

\journal{}

\begin{document}

\begin{frontmatter}

\title{Asymmetric volatility connectedness on forex markets\tnoteref{label1}\\}

\author[ies,utia]{Jozef Barun\'{\i}k\corref{cor2}}
\author[ies]{Ev\v{z}en Ko\v{c}enda}
\author[ies,utia]{Luk\'a\v{s} V\'acha}

\address[ies]{Institute of Economic Studies, Charles University, Opletalova 26, 110 00, Prague, Czech Republic}
\address[utia]{Institute of Information Theory and Automation, The Czech Academy of Sciences, Pod Vodarenskou Vezi 4, 182 00, Prague, Czech Republic}


\tnotetext[label1]{We are thankful for valuable comments from Lukas Menkhoff and presentation participants at Kyoto University. Support from GA\v{C}R grant No. 14-24129S is gratefully acknowledged. The paper was written while Ev\v{z}en Ko\v{c}enda was a Visiting Professor at the Institute of Economic Research, Kyoto University, whose hospitality is greatly appreciated. The usual disclaimer applies.}

\date{}

\begin{abstract}
{\small \noindent We show how bad and good volatility propagate through forex markets, i.e., we provide evidence for asymmetric volatility connectedness on forex markets. Using high-frequency, intra-day data of the most actively traded currencies over 2007 -- 2015 we document the dominating asymmetries in spillovers that are due to bad rather than good volatility. We also show that negative spillovers are chiefly tied to the dragging sovereign debt crisis in Europe while positive spillovers are correlated with the subprime crisis, different monetary policies among key world central banks, and developments on commodities markets. It seems that a combination of monetary and real-economy events is behind the net positive asymmetries in volatility spillovers, while fiscal factors are linked with net negative spillovers.}
\end{abstract}

\end{frontmatter}


\newpage
\section{Introduction}
A well documented, stylized fact of the asymmetric volatility phenomenon (AVP) indicates that volatility on financial markets is higher (lower) following market downturns (upturns).\footnote{See for example \cite{black1976,christie1982stochastic,pindyck1984,french1987expected}} However, AVP is not an isolated feature because volatility spills over across assets and markets quickly and its extent is captured by the volatility connectedness \citep{diebold2015financial}. While AVP has been studied intensively, asymmetries in volatility spillovers have not received enough attention despite the fact that volatility spillovers impact portfolio diversification strategies, portfolio management \citep{GarciaTsafack2011, aboura2014cross}, options and hedging strategies \citep{jayasinghe2008exchange,james2012handbook}. In this paper we do not analyz the AVP but investigate asymmetries in volatility spillovers. Recently, asymmetric volatility connectedness was documented among a set of U.S. stocks \citep{barunik2016asymmetric} and oil commodities \citep{barunik2015} but so far there is virtually no evidence related to forex markets. In this paper we generalize a quantification of asymmetric volatility connectedness and apply it on forex markets. The economic importance of our analysis rests in that we can learn in detail the dynamics of the asymmetries in volatility spillovers. Such assessment is impossible to learn from earlier work because there is no established procedure able to provide the same extent of detail and accuracy with which we could compare our results.

Our analysis is motivated by relevant questions arising with respect to spillovers in the forex markets. Do asymmetries in volatility spillovers exist among currencies? If they do, in what manner do they propagate? One currency might be prone to attract volatility spillovers in a manner different from other currency. Hence, is the extent and direction of spillover transmission among currencies uniform or dissimilar? And are the asymmetries in volatility spillovers and their directions uniform with regard to currencies, timing, and potential underlying factors, or do they exhibit differences?

The above questions are not trivial because the forex market differs from other financial markets in a number of ways. First, 24-hour operation across continents makes the forex market a truly global market with expansive information flow. Second, the forex market exhibits a very high degree of integration, especially for key currencies \citep{kitamura2010testing}. Third, the daily forex market turnover is in multiples of trading volumes on capital markets\citep{bis2013a}.\footnote{According to the latest Triennial Central Bank Survey issued by the Bank for International Settlements \citep[p.3]{bis2013a}, “trading in foreign exchange markets averaged \$5.3 trillion per day in April 2013. This is up from \$4.0 trillion in April 2010 and \$3.3 trillion in April 2007.” To contrast the above figures with trading volumes on capital markets, the global value of share trading in 2013 was \$55 trillion and represents a 12\% increase with respect to 2012 \cite[p.2]{wfe2014}. Still, with 251 trading days a year on average, daily share trading volume in 2013 represents about \$219 billion, a figure that is dwarfed by the turnover of the forex market.} Fourth, exchange rates of currency pairs are affected by monetary policies and interventions more than stocks and bonds. Notably, an increase or decrease in a differential between two (central bank) policy interest rates results (via monetary and economic channels) in a subsequent appreciation or depreciation of the specific currencies \citep{Taylor2001,devereux2003monetary,dick2015exchange}. The degree of uncertainty about monetary policies also affects exchange rate volatility and its spillovers. Fifth, central bank interventions often successfully impact the level and volatility of exchange rates, especially in emerging markets \citep{menkhoff2013foreign,fratzscher2015foreign}. Finally, it has been shown that the volatility connectedness of the forex market increased only mildly following the 2007 financial crisis and is also more stable when compared to other market segments such as trading stocks or bonds \citep[p.164]{diebold2015financial}.

Due to the above differences and to the unique features of the forex market, volatility spillovers among currencies might propagate and affect currencies' portfolios in less-than-intuitive ways. As \cite{Kanas2001} argues, positive and significant volatility spillovers may increase the nonsystematic risk that diminishes gains from international portfolio diversification -- this is even more important in light of the evidence that systematic volatility plays a dominant role in volatility spillovers among the world currencies \citep{greenwoodrisk}. In addition, \cite{Amonlirdviman2010} explicitly show that the asymmetry in the correlations of returns decreases the gains from international portfolio diversification. Based on this evidence it is reasonable to hypothesize that qualitative differences in shocks might produce qualitatively different volatility spillovers. In plain words, volatility due to positive or negative returns might induce differing volatility spillovers within a portfolio of currencies.

To the best of our knowledge there are almost no studies addressing the issue of asymmetries in foreign exchange volatility spillovers (asymmetric forex volatility connectedness). The exception is \cite{galagedera2012effect}, who model the interaction between returns and volatility in an autoregressive five-equation system and account for asymmetries in spillovers. They show that during the subprime crisis, depreciation of the U.S. dollar against the yen has a greater impact on U.S. dollar-yen volatility spillover than appreciation. On the other hand, the appreciation and depreciation of the U.S. dollar against the euro does not appear to have an asymmetric effect on euro-U.S. dollar volatility spillover. However, while we fully acknowledge the effort of this study, the methodological approach adopted imposes limits on its ability to capture the dynamics of asymmetries in volatility spillovers.

Connectedness measures based on network models seem to answer the need to improve the detection and measurement of spillovers along with their dynamics \citep{diebold2014network}. In their seminal work, \cite{diebold2009measuring} developed a volatility spillover index (the DY index) based on forecast error variance decompositions from vector autoregressions (VARs) to measure the extent of volatility transfer among markets. This methodology has been further improved in \cite{diebold2012better}, who used a generalized VAR framework in which forecast-error variance decompositions are invariant to variable ordering. The DY index is a versatile measure allowing dynamic quantification of numerous aspects of volatility spillovers. An important input to compute the DY index is realized variance that, however, does not allow accounting for asymmetries in volatility spillovers. On the other hand, the realized semivariances introduced by \cite{shephard2010measuring} enable one to isolate and capture negative and positive shocks to volatility and thus are ideally suited to interpreting qualitative differences in volatility spillovers.\footnote{The technique was quickly adopted in several recent contributions, see e.g. \cite{fenou2013,patton2014good,segal2015good}. Full details on the DY index and realized semivariances is provided in section \ref{sec:metodology}.}

\cite{barunik2016asymmetric} combine the ideas of both the DY index and realized semivariances and devise a way to measure asymmetries in volatility spillovers that are due to qualitatively different, positive or negative, returns. We modify their approach to better account for the transfer of spillovers on the forex market. Instead of using two separate $N$-dimensional VAR systems to measure asymmetries, we suggest a general framework where the negative and positive realized semivariances are in one system. Thus, we propose a $2N$-dimensional VAR resulting in a $2N\times 2N$ system of forecast variance error decompositions. The above modification results in versatile measure allowing dynamic quantification of asymmetric connectedness.\footnote{Full details of the formal exposition is provided in section \ref{sec:metodology}.} We then empirically apply our generalized framework on the forex data. For the purpose of verbal interpretation we adopt the terminology established in the literature \citep{patton2014good,segal2015good} to distinguish asymmetry in spillovers that originates due to qualitatively different uncertainty: bad uncertainty is defined as the volatility associated with negative innovations to quantities (e.g., output, returns) and good uncertainty as the volatility associated with positive shocks to these variables. We follow this terminology and label our spillovers as bad and good volatility spillovers (or negative and positive spillovers).

Hence, in our paper we provide two distinct contributions. First, we generalize the framework and modify the spillover asymmetry measure (SAM) introduced in \cite{barunik2016asymmetric} in order to isolate asymmetries in volatility spillovers among currencies on the forex market. Second, we then apply the method to analyzing asymmetries in volatility spillovers among major world currencies during specific periods of the global financial crisis and afterward. In doing so, we provide detailed results that are not available in any earlier study related to the researched topic. Specifically, we document the dominating asymmetries in spillovers that are due to bad rather than good volatility. We also show that negative spillovers are chiefly tied to the dragging sovereign debt crisis in Europe while positive spillovers are correlated with the subprime crisis, different monetary policies among key world central banks, and developments on commodities markets. It seems that a combination of monetary and real-economy events is behind the net positive asymmetries in volatility spillovers, while fiscal factors are linked with net negative spillovers.

The rest of the paper is organized in the following way. In Section \ref{sec:lit} we provide an overview of the literature related to forex volatility spillovers. In Section \ref{sec:metodology} we formally introduce the methodological approach and formulate testable hypotheses. Forex data are described in Section \ref{sec:Data} and in Section \ref{sec:results} we detail our results along with inferences and comments. Finally, conclusions are offered in Section \ref{sec:conclusion}.

\section{Literature review \label{sec:lit}}
Analyses of volatility spillovers date back to \cite{engle1990meteor}, who showed the existence of intra-day volatility spillovers on the forex market (meteor shower hypothesis) rather than being country-specific (heat wave hypothesis). Later, \cite{baillie1991intra} did not find enough evidence for systematic volatility spillovers among exchange rates while \cite{hong2001test} did find it, including directional spillovers from the former Deutsche mark to the Japanese yen. \cite{melvin2003global} used a non-parametric approach and analyzed the same pair of currencies across regions (Asia, Asia-Europe overlap, Europe, Europe-America overlap, America) and provided evidence of both intra- and inter-regional spillovers with intra-regional volatility spillovers being stronger. Similar evidence of volatility spillovers is given by \cite{cai2008informational}, who analyze spillovers in the euro-dollar and dollar-yen pairs across five trading regions. They find informational linkages to be statistically significant at both the own-region and inter-region levels, but volatility spillovers within a region dominate in terms of economic significance. \cite{kitamura2010testing} employs an MGARCH model, analyzes intra-day interdependence and volatility spillovers, and demonstrates that volatility spillovers from the euro significantly affect the Swiss franc and Japanese yen; the analysis is limited to the period July 2008 -- July 2009, though.

Network models analyzing connectedness have been gradually employed in the economic and financial literature but their application on forex markets is still limited. Some recent contributions provide quite specific results that are derived from the application of the DY index or build upon this concept. \cite[Chapter 6]{diebold2015financial} analyze the exchange rates of nine major currencies with respect to the U.S. dollar (USD) over 1999 to mid-2013. They show that forex market connectedness increased only mildly following the 2007 financial crisis: it exhibits numerous more and less pronounced cycles, but it is not linked to a business cycle. Directional volatility spillovers differ among currencies considerably. As both the U.S. dollar and the euro are the leading vehicle currencies of the global forex market, the EUR/USD exchange rate exhibits the highest volatility connectedness among all analyzed currencies.

\cite{greenwoodrisk} generalize the connectedness framework and analyze risk-return spillovers among the G10 currencies between 1999 and 2014 and find that spillover intensity is countercyclical and volatility spillovers across currencies increase during crisis times. Similarly, \cite{bubak2011volatility} document statistically significant intra-regional volatility spillovers among the European emerging foreign exchange markets and show that volatility spillovers tend to increase in periods characterized by market uncertainty, especially during the 2007 -- 2008 financial crisis. Further, \cite{mcmillan2010return} document the existence of volatility spillovers among the exchange rates of the U.S. dollar, British pound, and Japanese yen with respect to the euro and show dominating effects coming from the U.S. dollar. Finally, Antonakakis (2012) analyzes volatility spillovers among major currencies before and after the introduction of the euro and shows that the euro (Deutsche mark) is the dominant net transmitter of volatility, while the British pound is the dominant net receiver of volatility in both periods.

Among analyses that combine the assessment of volatility spillovers on the forex and other financial markets, the most frequent are those analyzing volatility interactions between the forex and stock markets. \cite{grobys2015volatility}, employing the DY index, finds very little evidence of volatility spillovers during quiet economic development but a high level of total volatility spillovers following periods of economic turbulence. A similar conclusion is found by \cite{do2015realized}, who also emphasize that it is important to account for the volatility spillover information transmission especially during the turbulent periods. Further, significant directional spillovers are identified between the forex and stock markets in several studies targeting developed and emerging markets \citep{do2016stock,andreou2013stock,kumar2013returns,Kanas2001} or specific countries or regions including the U.S. \citep{ito2015high}, Japan \citep{jayasinghe2008exchange}, China \citep{zhao2010dynamic}, the Middle East, and North Africa \citep{arfaoui2015return}.

Finally, some studies analyze interactions and volatility spillovers between the forex market and various segments of financial markets, such as stocks and bonds \citep{clements2015volatility}, commodities \citep{salisu2013modeling}, or stocks, bonds and commodities \citep{diebold2009measuring,duncan2013domestic,aboura2014cross,ghosh2014volatility}. However, the effects of asymmetries in volatility spillovers are analyzed in none of them.

\section{Measuring asymmetric volatility spillovers \label{sec:metodology}}
Seminal papers by \cite{diebold2009measuring} and \cite{diebold2012better}, along with other related studies, estimate volatility spillovers on daily (or weekly) high, low, opening, and closing prices. Estimators based on daily data offer, in general, good approximations of volatility. However, the low sampling frequency imposes some limitations. Having high-frequency data, we estimate volatility with convenient realized volatility estimators. Furthermore, to account for volatility spillover asymmetries, we follow \cite{barunik2015,barunik2016asymmetric}, who use the realized semivariance framework of \cite{shephard2010measuring}, which offers an interesting possibility to decompose volatility spillovers due to negative and positive returns. The quantification of asymmetric volatility spillovers with realized semivariances was first employed in \cite{barunik2015}, where the authors define measures using two separate VAR systems for negative and positive semi-variances. In this paper, to estimate asymmetric volatility spillovers, we define a more general approach with a single VAR system employing volatility spillovers from both negative and positive returns.

In this section, we first introduce the two existing concepts of total and directional spillovers from \cite{diebold2012better}, and then we describe a simple way to use realized semivarinces in order to capture asymmetric volatility spillovers. In order to keep our description on a general level, we will label variables as assets.

\subsection{Measuring volatility spillovers \label{sec:SI}}
The volatility spillover measure introduced by \cite{diebold2009measuring} is based on a forecast error variance decomposition from vector auto regressions (VARs). The forecast error variance decomposition traces how much of the $H$-step-ahead forecast error variance of a variable $i$ is due to innovations in another variable $j$, thus it provides an intuitive way to measure volatility spillovers. For $N$ assets, we consider an $N$-dimensional vector of realized volatilities, $\mathbf{RV_t} = (RV_{1t},\ldots,RV_{Nt})'$, to measure total volatility spillovers. In order to measure asymmetric volatility spillovers, we decompose daily volatility into negative (and positive) semivariances that provides a proxy for downside (and upside) risk. Using semivariances allows us to measure the spillovers from bad and good volatility and test whether they are transmitted in the same magnitude \citep{barunik2016asymmetric}. In this case we use a $2N$-dimensional vector, $\mathbf{RS_t} = (RS^{-}_{1t},\ldots,RS^{-}_{Nt},RS^{+}_{1t},\ldots,RS^{+}_{Nt})'$, consisting of positive and negative semivariances.

We start describing the procedure for the $N$-dimensional vector $\mathbf{RV_t} = (RV_{1t},\ldots,RV_{Nt})'$ and later extend the framework to accommodate realized semivariance.
Let us model the $N$-dimensional vector $\mathbf{RV_t}$ by a weakly stationary vector autoregression VAR($p$) as:
\begin{equation}
\label{RV}
\mathbf{RV_t} = \sum_{i=1}^p \mathbf{\Phi}_i \mathbf{RV}_{t-i}+ \boldsymbol{\epsilon}_t,
\end{equation}
where $\boldsymbol{\epsilon}_t\sim N(0,\mathbf{\Sigma}_{\epsilon})$ is a vector of $iid$ disturbances and $\mathbf{\Phi}_i$ denotes $p$ coefficient matrices. For the invertible VAR process, the moving average representation has the following form:
\begin{equation}
\mathbf{RV}_t = \sum_{i=0}^{\infty}\mathbf{\Psi}_{i}\boldsymbol{\epsilon}_{t-i}.
\end{equation}
The $N\times N$ matrices holding coefficients $\mathbf{\Psi}_i$ are obtained from the recursion $\mathbf{\Psi}_i = \sum_{j=1}^p\mathbf{\Phi}_j \mathbf{\Psi}_{i-j}$, where $\mathbf{\Psi}_0=\mathbf{I}_N$ and $\mathbf{\Psi}_i = 0$ for $i<0$. The moving average representation is convenient for describing the VAR system's dynamics since it allows disentangling the forecast errors. These are further used for the computation of the forecast error variances of each variable in the system, which are attributable to various system shocks. However, the methodology has its limitations as it relies on the Cholesky-factor identification of VARs. Thus, the resulting forecast variance decompositions can be dependent on variable ordering. Another important shortcoming is that it allows measuring total spillovers only. Therefore, \cite{diebold2012better} use the generalized VAR of \cite{koop1996impulse} and \cite{pesaran1998generalized} to obtain forecast error variance decompositions that are invariant to variable ordering in the VAR model and it also explicitly includes the possibility to measure directional volatility spillovers.\footnote{The generalized VAR allows for correlated shocks, hence the shocks to each variable are not orthogonalized.}

\subsubsection{Total spillovers\label{sec:tot}}
In order to define the total spillover index of \cite{diebold2012better}, we consider: (i) assets' own variance shares as fractions of the $H$-step-ahead error variances in forecasting the $i$th variable that are due to the assets' own shocks to $i$ for $i=1,\ldots,N$ and (ii) cross variance shares, or spillovers, as the fractions of the $H$-step-ahead error variances in forecasting the $i$th variable that are due to shocks to the $j$th variable, for $i,j=1,\ldots,N$, $i\ne j$. Then, the $H$-step-ahead generalized forecast error variance decomposition matrix $\Omega$ has the following elements for $H=1,2,\ldots$.
\begin{equation}
\omega_{ij}^H=\frac{\sigma_{jj}^{-1}\sum_{h=0}^{H-1}\left( \mathbf{e}'_i \mathbf{\Psi}_h \mathbf{\Sigma}_{\epsilon}\mathbf{e}_j \right)^2}{\sum_{h=0}^{H-1}\left( \mathbf{e}'_i \mathbf{\Psi}_h \mathbf{\Sigma}_{\epsilon}\mathbf{\Psi}'_h\mathbf{e}_i \right)}, \hspace{10mm} i,j=1,\ldots, N,
\end{equation}
where $\mathbf{\Psi}_h$ are moving average coefficients from the forecast at time $t$; $\mathbf{\Sigma}_{\epsilon}$ denotes the variance matrix for the error vector, $\boldsymbol{\epsilon}_t$; $\sigma_{jj}$ is the standard deviation of the error term for the $j$th equation; $\mathbf{e}_i$ and $\mathbf{e}_j$ are the selection vectors, with one as the $i$th or $j$th element and zero otherwise.

As the shocks are not necessarily orthogonal in the generalized VAR framework, the sum of the elements in each row of the variance decomposition table is not equal to one. Thus, we need to normalize each element by the row sum as:
 \begin{equation}
 \widetilde{\omega}_{ij}^H = \frac{\omega_{ij}^H}{\sum_{j=1}^N \omega_{ij}^H}.
 \end{equation}
\cite{diebold2012better} then define the total spillover index as the contribution of spillovers from volatility shocks across variables in the system to the total forecast error variance, hence:
\begin{equation}
\label{stot}
\mathcal{S}^H=100\times \frac{1}{N} \sum_{\substack{i,j=1\\ i\ne j}}^N\widetilde{\omega}_{ij}^H.
\end{equation}
Note that $\sum_{j=1}^N \widetilde{\omega}_{ij}^H=1$ and $\sum_{i,j=1}^N \widetilde{\omega}_{ij}^H=N$. Hence, the contributions of spillovers from volatility shocks are normalized by the total forecast error variance. To capture the spillover dynamics, we use a 200-day rolling window running from point $t-199$ to point $t$. Further, we set the forecast horizon $H=10$, and a VAR lag length of 2.\footnote{In addition, we constructed the spillover index with rolling windows of 150 and 100 days to check the robustness of our results. We have also experimented with different $h$ values, and we find that the results do not materially change and are robust with respect to the window and horizon selection. The VAR lag length was chosen based on AIC to produce the most parsimonious model.}

\subsubsection{Directional spillovers \label{sec:dir}}
The total volatility spillover index indicates how shocks to volatility spill over all the assets. However, with the generalized VAR framework, we are able to identify directional spillovers using the normalized elements of the generalized variance decomposition matrix \citep{diebold2012better}. The directional spillovers are important, as they allow us to uncover the spillover transmission mechanism disentangling the total spillovers to those coming from or to a particular asset in the system.

Following \cite{diebold2012better} we measure the directional spillovers received by asset $i$ from all other assets $j$:
\begin{equation}
\mathcal{S}_{N,i\leftarrow\bullet}^H=100\times \frac{1}{N} \sum_{\substack{j=1\\ i\ne j}}^N\widetilde{\omega}_{ij}^H,
\end{equation}
i.e., we sum all numbers in rows $i$, except the terms on a diagonal that correspond to the impact of asset $i$ on itself. The $N$ in the subscript denotes the use of an $N$-dimensional VAR. Conversely, the directional spillovers transmitted by asset $i$ to all other assets $j$ can be measured as the sum of the numbers in the column for the specific asset, again except the diagonal term:
\begin{equation}
\mathcal{S}_{N,i\rightarrow\bullet }^H=100\times \frac{1}{N} \sum_{\substack{j=1\\ i\ne j}}^N\widetilde{\omega}_{ji}^H.
\end{equation}

As we now have complete quantification of how much an asset receives (transmits), denoted as the direction from (to), we can compute how much each asset contributes to the volatility in other assets in net terms. The net directional volatility spillover from asset $i$ to all other assets $j$ is defined as the difference between gross volatility shocks transmitted to and received from all other assets:
\begin{equation}
\mathcal{S}^H_{N,i}=\mathcal{S}_{N,i\rightarrow\bullet }^H-\mathcal{S}_{N,i\leftarrow\bullet}^H.
\end{equation}

\subsection{Measuring asymmetric spillovers \label{sec:MAS}}
Using the advantage of high-frequency data, we can track the asymmetric behavior of volatility spillovers. In particular, we are able to distinguish spillovers from volatility due to negative returns and positive returns (bad and good volatility). Further, we are also able to distinguish directional volatility spillovers (in the direction TO) due to negative returns and positive returns.\footnote{We do not estimate directional volatility spillovers FROM as it is difficult to interpret these in the $2N\times 2N$ spillover matrix setting.} Following \cite{barunik2015} and \cite{barunik2016asymmetric}, we first disentangle daily realized volatility into negative and positive daily realized semivariances (for more details see the Appendix). The semivariances allow us to estimate volatility spillovers due to bad or good volatility and quantify asymmetries in spillovers. For $N$ assets, \cite{barunik2015} use two separate $N$-dimensional VAR systems to measure the asymmetries. In this paper, we propose a more general framework where negative and positive realized semivariances are employed in a single VAR. Thus, we estimate a $2N$-dimensional VAR, resulting in $2N\times 2N$ system of forecast variance error decompositions.

As our empirical analysis, based on the described methodological approach, employs forex data, we will use the term currency (instead of asset) from now on. In order to obtain asymmetric volatility spillovers for $N$ currencies, we construct a VAR model (Eq. \ref{RV}), but we replace the vector of realized volatilities $\mathbf{RV_t} = (RV_{1t},\ldots,RV_{Nt})'$ with the $2N$ dimensional vector of negative and positive semivariances $\mathbf{RS_t} = (RS^{-}_{1t},\ldots,RS^{-}_{Nt},RS^{+}_{1t},\ldots,RS^{+}_{Nt})'$. Then the elements of $2N\times 2N$ $H$-step-ahead generalized forecast error variance decomposition matrix $\Omega$ has the form:
\begin{equation}
\omega_{ij}^H=\frac{\sigma_{jj}^{-1}\sum_{h=0}^{H-1}\left( \mathbf{e}'_i \mathbf{\Psi}_h \mathbf{\Sigma}_{\epsilon}\mathbf{e}_j \right)^2}{\sum_{h=0}^{H-1}\left( \mathbf{e}'_i \mathbf{\Psi}_h \mathbf{\Sigma}_{\epsilon}\mathbf{\Psi}'_h\mathbf{e}_i \right)}, \hspace{10mm} i,j=1,\ldots, 2N,
\end{equation}
where $\mathbf{\Psi}_h$ denotes the moving average coefficient matrix from the forecast at time $t$; $\mathbf{\Sigma}_{\epsilon}$ is the variance matrix for the error vector $\boldsymbol{\epsilon}_t$; $\sigma_{jj}$ is the standard deviation of the error term for the $j$th equation; $\mathbf{e}_i$ and $\mathbf{e}_j$ are the selection vectors, with one as the $i$th or $j$th element and zero otherwise.

\subsubsection{Directional spillover asymmetry measure}
Standard directional spillovers give us an important insight about the volatility spillovers' structure among the studied currencies. However, we may benefit from realized semivariances to obtain more precise information about spillover behavior by defining a directional spillover asymmetry measure. The asymmetry is defined as the difference between the directional volatility spillover coming from a positive or negative semivariance. The standard directional spillovers are defined in Section \ref{sec:dir} for both directions, i.e. FROM and TO. However, in the case of asymmetry we define only the direction TO as its interpretation is straightforward in the $2N\times 2N$ spillover matrix setting while the interpretation of FROM is quite vague. Specifically, we define directional asymmetries in volatility spillovers coming from a specific currency TO the rest of the currencies under research.

In Table \ref{tab:spillsetting} we show the elements of the $2N\times 2N$ $H$-step-ahead generalized forecast error variance decomposition matrix $\Omega$ for a specific case of the six currencies we analyze (at this moment we refrain from introducing the currencies and leave details to Section \ref{sec:Data}). To compute directional spillovers, in the direction TO, we sum the corresponding column of the $2N \times 2N$ spillover matrix (Table \ref{tab:spillsetting}) excluding the own share on the main diagonal, $i\neq j$, and two diagonals in the $N \times N$ block sub-matrices (lower left and upper right), i.e., $\vert i-j\vert \neq N/2$. All excluded numbers are highlighted in bold, hence for every column we sum $2N-2$ numbers. We define directional spillover from a currency $i$ to all other currencies as:
\begin{equation}
\mathcal{S}^H_{2N,i\rightarrow\bullet }=100\times \frac{1}{2N} \sum_{\substack{i=1, i\ne j\\ \vert i-j\vert \neq N/2}}^{2N}\widetilde{\omega}_{j,i}^H, \hspace{10mm} i,j=1,\ldots, 2N.
\end{equation}
Based on directional spillovers, we can now introduce the net asymmetric directional spillovers that measure how shocks from bad and good volatility to one currency affect the volatility of all other currencies. Let us define the directional spillover asymmetry measure as the difference of the response to a shock from bad or good volatility from currency $i$ to other currencies. Thus, for currency $i$ we subtract the effect of the $(N+i)$-th column of a spillover matrix from the effect of the $i$-th column, i.e.,
\begin{equation}
\label{samd}
\mathcal{SAM}^H_{2N,i\rightarrow\bullet} =  \mathcal{S}^H_{2N,i\rightarrow\bullet } - \mathcal{S}^H_{2N,(i+N)\rightarrow\bullet }, \hspace{10mm} i,\ldots, N.
\end{equation}
If the $\mathcal{SAM}^H_{N,i\rightarrow\bullet}$ is negative (positive), then we observe a stronger effect of bad (good) volatility of currency $i$ to other currencies. Again, to capture the time-varying nature of spillovers, we use a 200-day moving window running from point $t-199$ to point $t$.

\subsubsection{Spillover asymmetry measure\label{samsec}}
While the spillover asymmetry measure defined by Equation (\ref{samd}) gives us detailed information about the extent of asymmetry only for one currency, we can now define a measure that describes the volatility spillover asymmetry for the whole system (portfolio) of currencies. The idea of a spillover asymmetry measure ($\mathcal{SAM}$) was introduced in \cite{barunik2015} -- however, we extend their approach by using all available volatility spillovers in one $2N$-dimensional VAR model.\footnote{The subscript $2N$ in the spillover asymmetric measure and the directional measures denotes that a $2N$-dimensional VAR model was used for spillover computation.} We define the spillover asymmetry measure with an $H$-step-ahead forecast at time $t$, $\mathcal{SAM}^H_{2N}$, as a difference between volatility spillovers due to negative and positive returns for all currencies $N$:
\begin{equation}
\label{eq:sam}
\mathcal{SAM}^H_{2N}=\sum_{i=1}^{N} \mathcal{S}^H_{2N,i\rightarrow\bullet } - \sum_{i=N+1}^{2N} \mathcal{S}^H_{2N,i\rightarrow\bullet }.
\end{equation}
The $\mathcal{SAM}^H_{2N}$ help us to better understand the behavior of volatility spillovers for a given portfolio of assets. In case there is no spillover asymmetry, spillovers coming from $RS^-$ and $RS^+$ are equal, thus $\mathcal{SAM}^H_{2N}$ takes the value of zero. However, when $\mathcal{SAM}^H_{2N}$ is negative (positive), spillovers coming from $RS^-$ are larger (smaller) than those from $RS^+$. In order to test the null hypothesis of symmetric connectedness, we use bootstrap confidence intervals constructed as described by \cite{barunik2016asymmetric}.

\subsection{Hypotheses}
The previous definitions of $\mathcal{SAM}$ and the directional $\mathcal{SAM}$ (D -- $\mathcal{SAM}$) help us to better understand the behavior of volatility spillovers for a given portfolio of currencies. In case there is no spillover asymmetry, spillovers coming from $RS^-$ and $RS^+$ are equal and the $\mathcal{SAM}$ and D -- $\mathcal{SAM}$ take the value of zero. However, when the $\mathcal{SAM}$ and D -- $\mathcal{SAM}$ are negative (positive), spillovers coming from $RS^-$ are larger (smaller) than those from $RS^+$. This pattern would then clearly indicate the existence and extent of asymmetries in volatility spillovers. Following our exposition in Section \ref{sec:MAS}, we formulate several testable hypotheses of symmetric connectedness to test for the presence of potential asymmetries in volatility spillovers (asymmetric volatility connectedness) among currencies.

Hypothesis 1: Volatility spillovers in the portfolio of currencies do not exhibit asymmetries.
Formally, Hypothesis 1 is formulated as:
$$\begin{array}{ccccccc}
\mathcal{H}^1_0: &&\mathcal{SAM}^H_{2N} = 0& \text{against} & \mathcal{H}_A^1: && \mathcal{SAM}^H_{2N} \ne 0.\\
\end{array}$$

Hypothesis 2: No directional volatility spillovers coming from either $RS^-$ or $RS^+$ are transmitted from one currency to the rest of the currencies in a portfolio. Formally, Hypothesis 2 is formulated as:
$$\begin{array}{ccccccc}
\mathcal{H}^2_0: && \mathcal{S}^H_{2N,i\rightarrow\bullet } = 0 & \text{against} & \mathcal{H}_A^2: && \mathcal{S}^H_{2N,i\rightarrow\bullet } \ne 0  \hspace{4mm} (i=1,\ldots, 2N).\\
\end{array}$$

Hypothesis 3: Volatility spillovers transmitted from one currency do not exhibit an asymmetric impact on the volatility of the other currencies in portfolio.
Formally, the Hypothesis 3 is formulated as:
$$\begin{array}{ccccccc}
\mathcal{H}^3_0: &&\mathcal{SAM}^H_{2N,i\rightarrow\bullet} = 0& \text{against} & \mathcal{H}_A^3: && \mathcal{SAM}^H_{2N,i\rightarrow\bullet} \ne 0.\\
\end{array}$$

Rejecting a null hypothesis means that bad and good volatility does matter for spillover transmission in terms of magnitude as well as direction. Moreover, we assume that the values of the volatility spillover indices differ over time. To capture the time-varying nature of the potential asymmetries, we compute the indices using a 200-day moving window that runs from point $t-199$ to point $t$; more details were provided in Section \ref{sec:tot}. In order to test the null hypotheses of symmetric connectedness, we use bootstrap confidence intervals constructed as described by \cite{barunik2016asymmetric}.

\section{Data and dynamics \label{sec:Data}}
In this paper we compute volatility spillovers measures on the foreign exchange future contracts of six currencies over the period from January 2007 to December 2015. The currencies are the Australian dollar (AUD), British pound (GBP), Canadian dollar (CAD), euro (EUR), Japanese yen (JPY), and Swiss franc (CHF). All these currency contracts are quoted against the U.S. dollar (USD) and this is a typical approach in the forex literature (any potential domestic (U.S.) shocks are integrated into all currency contracts). The currencies under research constitute a group of the most actively traded currencies globally \citep{bis2013a,antonakakis2012exchange} and this is the reason for our choice: we aim at analyzing asymmetric connectedness among the currencies that constitute two thirds of the the global forex turnover by currency pair \citep{bis2013a}; we do not pursue assessment of less traded currencies at the moment.

The foreign exchange future contracts are traded on the Chicago Mercantile Exchange (CME) on a nearly 24-hour basis and transactions are recorded in Chicago time (CST). Trading activity starts at 5:00 pm CST and ends at 4:00 pm CST. To exclude potential jumps due to the one-hour gap in trading, we redefine the day in accordance with the electronic trading system. Furthermore, we eliminate transactions executed on Saturdays and Sundays, U.S. federal holidays, December 24 to 26, and December 31 to January 2, because of the low activity on these days, which could lead to estimation bias. The data are available from Tick Data, Inc.\footnote{http://www.tickdata.com/}

\begin{figure}[h]
   \centering
   \includegraphics[width=6in]{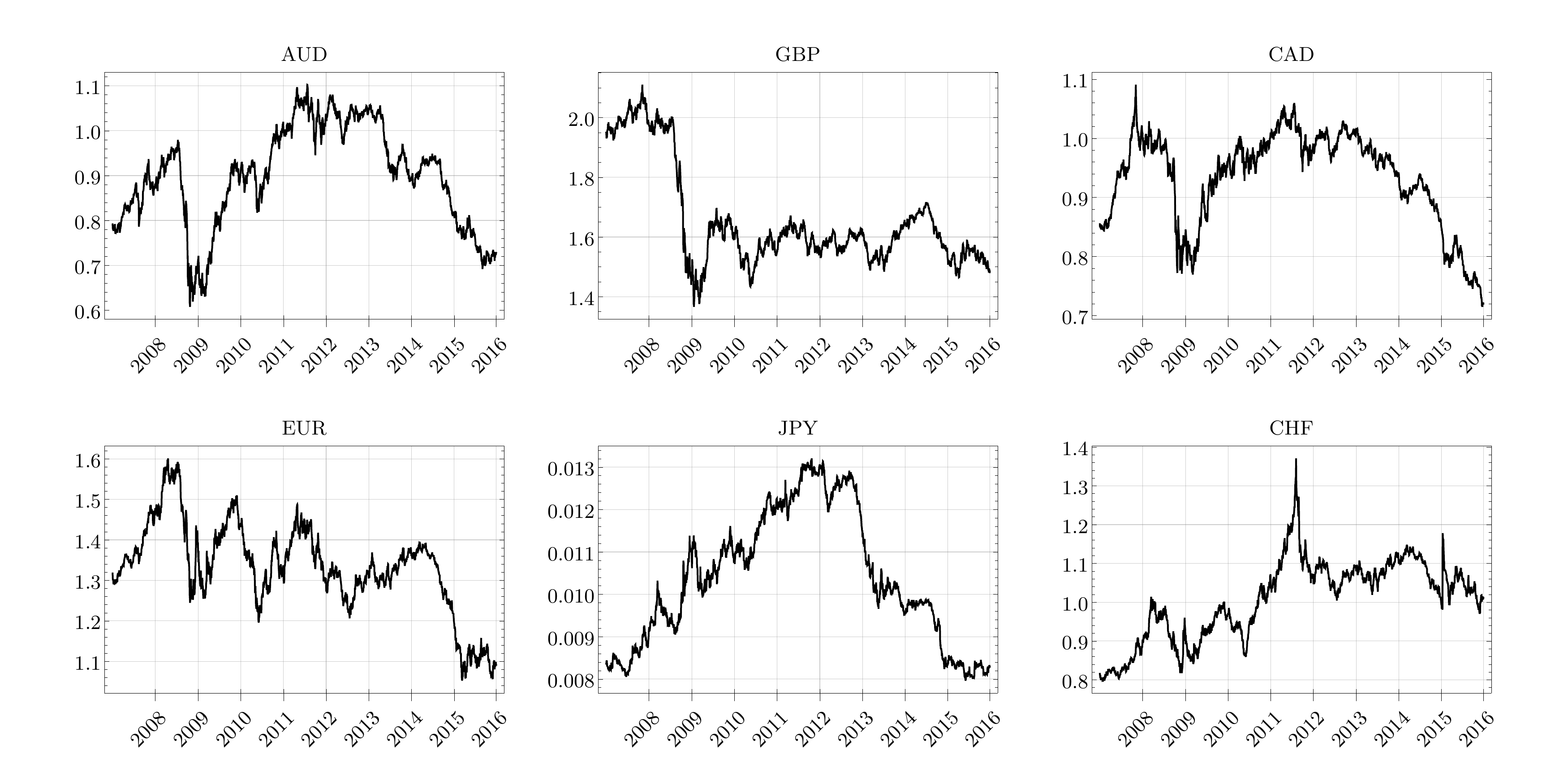}
   \caption{Foreign exchange future contracts of AUD, GBP, CAD, EUR, JPY and CHF quoted in the unit value of US dollars}
   \label{Fig1}
\end{figure}

In Figure \ref{Fig1} we plot the exchange rates of all six currencies (EUR, JPY, GBP, AUD, CHF, CAD). Each plot is labelled by the three-letter international code of the specific currency and exhibits the dynamics of the currency’s price in terms of the U.S. dollar over the sample period. The dynamics of the exchange rates is remarkably different and only two commodity currencies (AUD and CAD) share an overall common pattern. Still, all six currencies exhibit depreciation with respect to the USD following the 2008 global financial crisis (GFC) -- the extent differs and the Japanese and Swiss currencies show the least GFC-related depreciation. The remarkably stable path of the GBP from 2009 is in contrast to the post-GFC appreciation of other currencies, followed by a depreciation after 2012 (AUD, CAD, JPY). On the other hand, the euro to U.S. dollar exchange rate exhibits a series of ups and downs related to various major events among which the most important are the rounds of quantitative easing performed by the Fed between 2009 and 2014, and the key part of the EU debt crisis (2010 -- 2011). The Swiss franc shows a prominent wave of appreciation in 2011 and a subsequent depreciation after the managed float regime was given up by the Swiss National Bank.

\section{Results \label{sec:results}}

\subsection{Total connectedness and economic conditions}
\begin{figure}[h]
   \centering
   \includegraphics[width=6in]{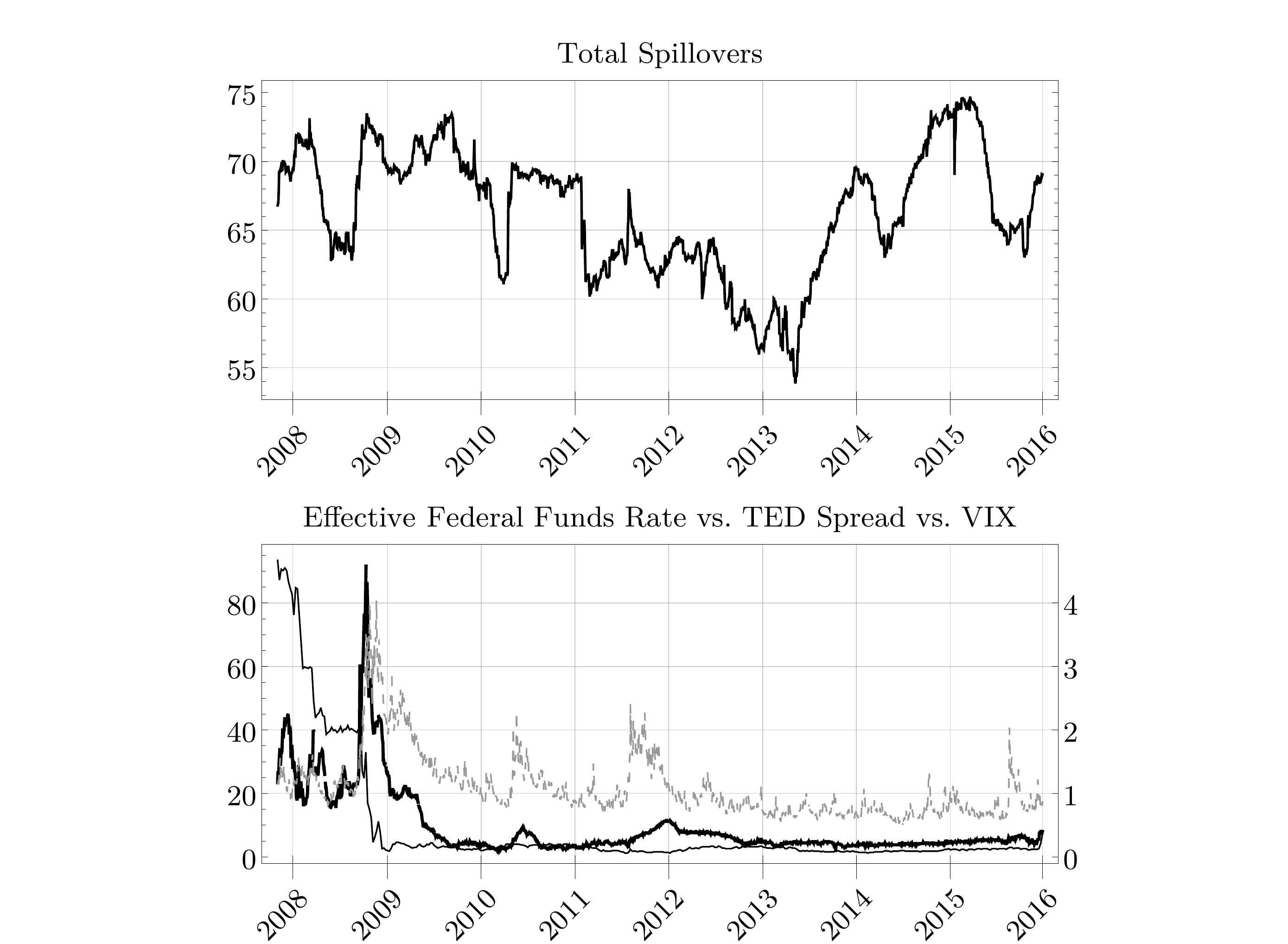}
   \caption{Upper panel: the total volatility spillovers of six currencies, Lower panel: the Federal funds rate, the TED and the VIX.}
   \label{Fig2}
\end{figure}

In Figure \ref{Fig2} (upper panel), we show the total connectedness among the six currency pairs. The total forex volatility spillovers measure is calculated based on \cite{ diebold2012better}: the connectedness is quite high during the GFC period, until 2010, and then in 2014. The total connectedness values of 65\% and above during the 2008 -- 2010 period are comparable to those found in \cite{diebold2015financial}. The plot exhibits a distinctive structural change in total connectedness among the six currencies under research: initial relatively stable and high connectedness (interrupted by a short drop during 2009) decreases gradually after 2010 but then in 2013 begins to rise sharply, surpassing in 2015 the original levels from the GFC period. The period is marked by two distinctive phenomena. One is the difference between monetary policies among the Fed, ECB, and Bank of Japan. While the Fed stopped the quantitative easing (QE) policy in 2014, the ECB was beginning to pursue it and the Bank of Japan was already active in pursuing this policy. From 2013 the policy differences affected the capital flows and carry-trade operations so that the U.S. dollar began to appreciate against the euro and yen. At the same time, falling commodity prices exerted downward pressure on inflation and interest rates. This course affects most of the currencies in our sample as commodities are quoted in vehicle currencies (USD, EUR, JPY) and interest rate cuts occurred for commodity currencies (AUD, CAD), diminishing their appeal for carry-trade activities. Hence, the increased volatility and spillovers among currencies from 2013 on are to be found in combined effects chiefly rooted in monetary steps.

In the lower panel of Figure \ref{Fig2}, we relate the total connectedness to economic conditions represented by the plots of three indicators: the Federal funds rate, the TED, and the VIX. Unfortunately, for most of the period under research the Federal funds rate is near the zero lower bound and this precludes assessing a link between the total connectedness and U.S. economic development.\footnote{For the earlier period, \cite{greenwoodrisk} document a negative correlation between the Federal funds rate and forex spillovers. The evidence is suggestive of the potential that the U.S. dollar drives much of the forex market dynamics \citep{lustig2011common}.} Further, we compare total connectedness and the TED. Both measures share maximum values in 2008 in association with the fall of Lehman Brothers and other GFC-related events. Then the TED decreases rapidly as the Fed began to lend money and to guarantee interbank lending. Spillovers start to decrease after 2010 as well. The pattern in movements of both measures indicates that forex spillovers seem to strengthen during a period of low liquidity on the market. Finally, we observe several instances when total connectedness increases along with spikes in the VIX in 2008, 2010, 2011, and 2015. Our interpretation is that forex spillovers tend to build up during periods of financial distress.

\subsection{Directional spillovers}\label{sec:ds}
We now turn to a more detailed analysis of spillovers among specific currencies. The total volatility connectedness (the upper panel of Figure \ref{Fig2}) exhibits the extent of volatility spillovers for all six currencies. Following \cite{diebold2012better} we are able to compute directional spillovers and show how volatility from a specific currency transmits to other currencies in our sample (``contribution TO"). Similarly we are also able to show the opposite link of the extent of spillovers going from other currencies to a specific currency (``contribution FROM"). The condensed information on the extent of such directional spillovers is presented in Table \ref{tab:spills}. The information presented within the table shows in aggregate form the differences in how specific currencies transmit and receive spillovers. The most important directional spillovers are detected between commodity currencies (AUD, CAD) and between the pairs EUR-CHF and EUR-GBP. However, these differences are highly aggregated and do not illustrate the evolution over time.

\begin{figure}[!ht]
   \centering
   \includegraphics[width=6in]{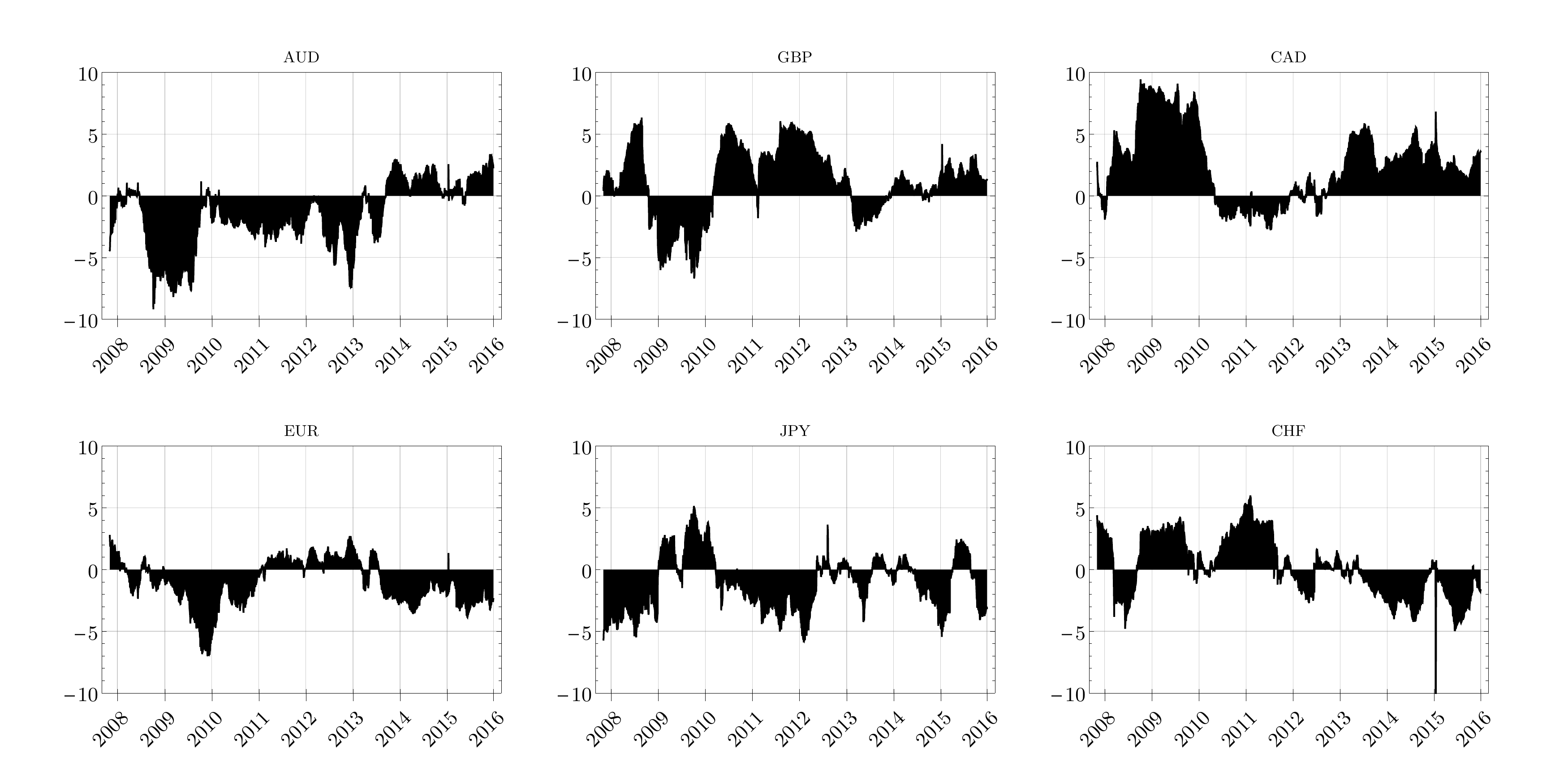}
   \caption{Net directional spillovers}
   \label{Fig3}
\end{figure}

Therefore, we compute the net effect of the directional spillovers: a difference between ``contribution TO" and ``contribution FROM" that we plot in Figure \ref{Fig3}, where the most interesting patterns emerge. The positive domain contains net spillovers that a currency transmits to other currencies and we say that a currency is a ``spillover giver". The net spillovers in the negative domain then represent the situation when a specific currency receives net volatility spillovers from others: in this case the currency is said to be a ``spillover receiver".

Figure \ref{Fig3} offers interesting insights based on the dynamic patterns that show each currency's net position in terms of the volatility spillovers it receives or transmits. One might hypothesize that the extent of spillover transmission among currencies is uniform. However, the evidence shown in Figure \ref{Fig3} shows quite the opposite. Both commodity currencies (AUD and CAD) can be characterized by opposite extreme net positions: AUD is a net volatility spillover receiver and CAD is a spillover giver; short periods when low net spillovers are in the opposite domains are exceptions. Exactly the opposite pattern is detected with JPY that clearly receives more spillovers during most of the researched period and transmits them moderately after the GFC began to subside. This behavior might be connected to the known intervention practice of the Bank of Japan. \cite{chortareas2013volatility} find that the Bank of Japan interventions in the USD/JPY exchange rate decrease (only in a short term of less than five hours and in a discontinuous pattern) the daily volatility of the USD/JPY rate. This suggests that the interventions can decrease volatility in the short run. The finding is in line with our results because during 2000 JPY behaved as a spillover giver (increased volatility) but during the rest of our period it was mostly a spillover receiver as its own volatility diminished relative to the rest of the currencies.

The rest of the currencies in Figure \ref{Fig3} alternate between being givers or receivers, depending on the time. Still, GBP could be described as being a more spillover-giving currency because it receives non-marginal net spillovers only during 2009, marking the financial crisis aftermath. EUR receives more net spillovers as the European sovereign debt crisis builds up and then from 2013 on. Despite being rather a spillover receiver, EUR seems to be the calmest currency as the net directional spillovers are quite low. The results for GBP and EUR are in line with those presented by \cite{antonakakis2012exchange} for the period 2000 -- 2012, who finds GBP and EUR to be the dominant net transmitter and net receiver of volatility, respectively.\footnote{\cite{antonakakis2012exchange} employs the DY spillover index approach and shows that the Deutsch mark (euro) is the dominant net transmitter of volatility, while the British pound is the dominant net receiver of volatility both before and after the introduction of the euro. The exchange rates in \cite{antonakakis2012exchange} are defined as the number of Deutsch mark/euro/GBP units per one USD. Hence, in this footnote we transposed his original interpretation of transmitter/receiver to correspond with our analysis because we define the exchange rate as the number of U.S. dollars per one unit of a specific currency.} The most balanced currency in terms of net spillovers is CHF where the spillover giver/receiver positions alternate quite often.

Finally, since we employ data origination in one market (the Chicago Mercantile Exchange), we are unable to test the meteor-shower hypothesis of \cite{engle1990meteor}. Instead, the above evidence on directional spillovers among currencies suggests the presence of heat-wave volatility clustering as the values of spillovers indicate that substantial spillovers are transmitted among currencies within a specific market. Thus, our results are also in line with those presented by \cite{melvin2003global} and \cite{cai2008informational}.

The above results do not involve asymmetries in volatility spillovers but they confirm earlier findings in the literature. This validation is important for our work in terms of accuracy because our extension of the Diebold-Yilmaz methodology provides assessment of the asymmetries in volatility connectedness, results of which we present below.

\subsection{Asymmetries in volatility spillovers}
So far we have shown evidence based on spillovers that did not account for asymmetries. Now, we will employ the realized semivariances to separate qualitatively different shocks to volatility. Details on the computation of realized semivarinces are described in \ref{sec:semi}. In short, negative realized semivariance ($RS^-$) isolates negative shocks to volatility or, in other words, $RS^-$ allows capturing volatility due to negative changes (returns) in an exchange rate. The opposite is true for positive realized semivariance ($RS^+$). The descriptive statistics of realized semivariances are reported in Table \ref{tab:descr}. The similarity in the values of the first two moments of the positive and negative semivariances hints at the similarity of both types of volatility measures. However, such similarity is misleading because differences in skewness and kurtosis (including minimum and maximum values) suggest that realized semivariances do not need to be similar after all, especially when we account for their dynamics.

\begin{figure}[!ht]
   \centering
   \includegraphics[width=5in]{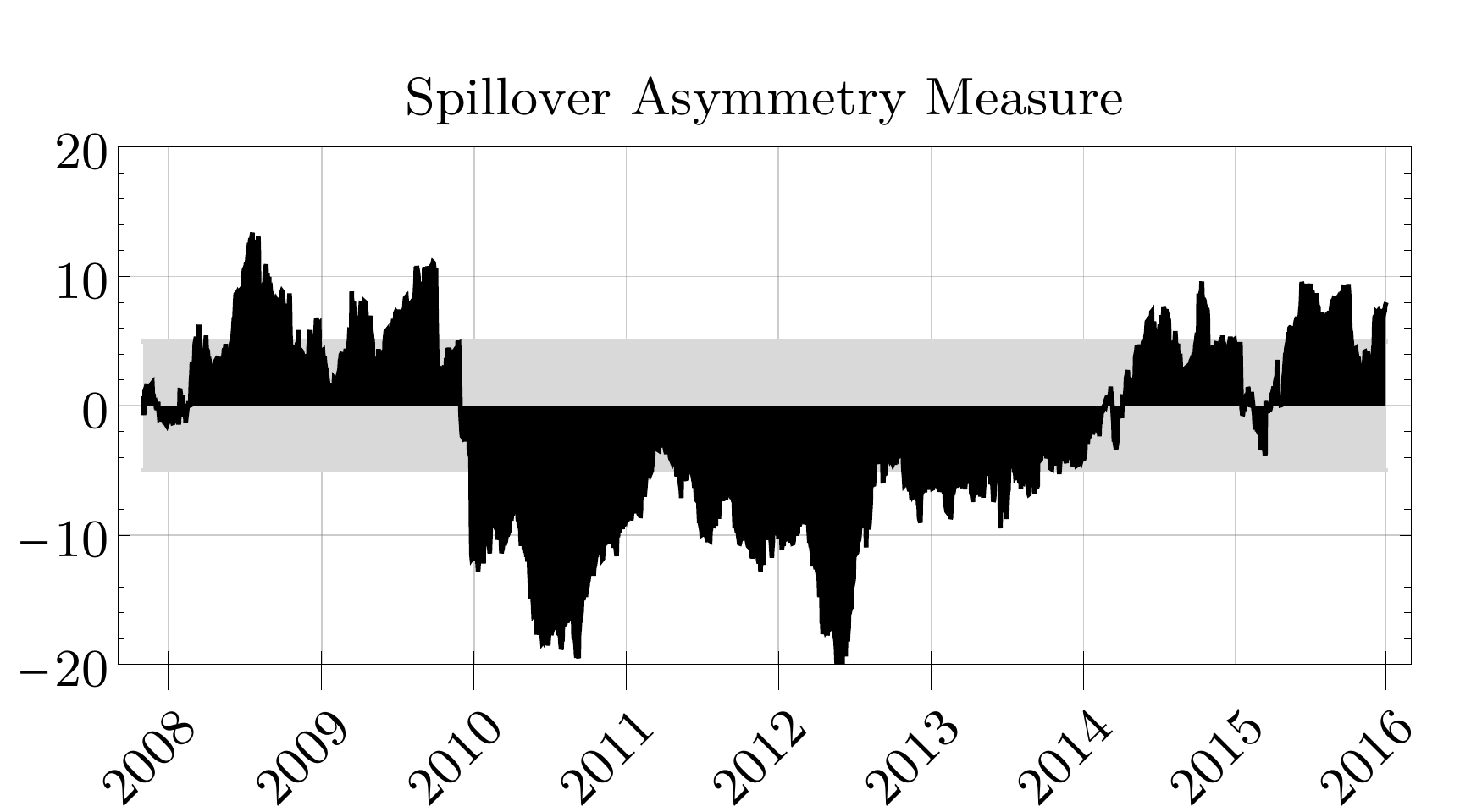}
   \caption{Spillover asymmetry measure - $\mathcal{SAM}$. Shaded band represents a 95\% confidence interval based on bootstrap}
   \label{Fig4}
\end{figure}

In contrast to Table \ref{tab:spills}, Figure \ref{Fig4} offers entirely new insights. It is the plot of the spillover asymmetry measure ($\mathcal{SAM}$) computed as the difference between the spillover indices for all six currency pairs where inputs are realized semivariances as in specification (\ref{eq:sam}), whose descriptive statistics are presented in Table \ref{tab:spills}. The volatility associated with negative (positive) innovations to returns has been termed as bad (good) volatility \citep{patton2014good,segal2015good}. We follow this terminology and label spillovers in Figure \ref{Fig4} as bad and good volatility spillovers (or simply negative and positive spillovers).

The plot of $\mathcal{SAM}$ in Figure \ref{Fig4} exhibits a similarly broken pattern as the total connectedness measure in Figure \ref{Fig2}, upper panel. However, a qualitatively new picture emerges. Asymmetries due to positive shocks measured with $RS^+$ are plotted in the positive domain and dominate the early and late periods of our sample (2008 -- 2009; 2014 -- 2015). On the other hand, during 2010 -- 2013 the asymmetries due to negative shocks measured with the $RS^-$ are plotted in the negative domain and dominate not only in their length but also in terms of their magnitude. Based on the evidence in Figure \ref{Fig4} we are able to reject Hypothesis 1 as the volatility spillovers in the portfolio of currencies exhibit distinctive asymmetries.

Further, the evidence suggests that different types of event are dominated by different types of spillover. The period of the global financial crisis (2007 -- 2009) that emerged in the U.S. is characterized by good volatility spillovers. This indicates that positive shocks dominated negative ones. In other words, asymmetries in volatility spillovers during the GFC were grounded chiefly in the good volatility of the currencies’ values with respect to the U.S. dollar. The period marked by the European sovereign debt crisis that fully unfolded in 2010 offers a different view. The asymmetries are more pronounced and bad volatility spillovers clearly dominate the period 2010-2013. The largest values mark the 2010 Greek fiscal crisis and in 2012 the combined major effects of the Greek vote against the austerity plan and Spain’s troubled situation that forced it to launch a rescue plan for its banking sector \citep{brei2013rescue}.

Besides the key events described above, there were other factors as well. The largest asymmetries due to negative shocks occurring in 2010 also reflect the development in the commodities’ markets: rising prices and the progressive financialization of commodities \citep{cheng2013financialization}. The pattern also correlates well with the improvement of the U.S. labor market and the development in emerging markets and China that are naturally paired with the development of the commodities as well. Large asymmetries in 2011 -- 2012 reflect further improvement in commodities markets until they burst. High asymmetries due to positive shocks in 2014 and later on should be paired with two major events. One, dramatically falling prices in commodities markets that resulted in interest-rate cuts by many central banks. Two, a prominent divide between the monetary policies of the Fed and its major counterparts (ECB, Bank of Japan) because international markets are quite sensitive to the Fed’s monetary policy as U.S. treasury securities dominate in global markets \citep[p. 32]{siklos2017}.

In terms of the interpretation related to asymmetries we assume that outbursts of good and bad volatilities of a specific currency spill over and increase the volatility of other currencies. The reasoning behind this assumption is that we study exchange rates involving seven currencies (USD, EUR, JPY, GBP, AUD, CHF, CAD) that account for almost 90\% of the global foreign exchange market turnover; further, the six highly traded currency pairs (with respect to the USD) based on the seven currencies that we study amount to two thirds of total global forex trading \citep{bis2013a}.\footnote{According to \cite[pp.10-11]{bis2013a}, the currency distribution of global foreign exchange market turnover is dominated by seven currencies (USD, EUR, JPY, GBP, AUD, CHF, CAD) that account for 173.6\% of the global forex market turnover (out of 200\% - the sum of the percentage shares of individual currencies totals 200\% instead of 100\% because two currencies are involved in each transaction). Further, the six currency pairs (USD/EUR, USD/JPY, USD/GBP, USD/AUD, USD/CAD, USD/CHF) amount to 65.1\% of the global foreign exchange market turnover by currency pair.} Since most of the trades in the currency markets are speculative in nature, the currencies in our sample can be considered substitutes.\footnote{The financial education website Investopedia states that “day-to-day corporate needs comprise only about 20\% of the market volume. Fully 80\% of trades in the currency market are speculative in nature (http://www.investopedia.com/articles/forex/06/sevenfxfaqs.asp; retrieved on March 10, 2016). The data provided by \cite[p.6]{bis2013a} do not provide a direct estimate of speculative trading but allow an indirect inference via foreign exchange market turnover by counterparty that is proportionally divided among non-financial customers (9\%), reporting dealers (39\%), and other financial institutions (53\%). Further, in terms of the instruments, “FX swaps were the most actively traded instruments in April 2013, at \$2.2 trillion per day, followed by spot trading at \$2.0 trillion” \citep[p.3]{bis2013a}. Hence, the figures also support the major role of the forex speculative trading.} Hence, volatility spillovers from one currency are assumed to directly impact the volatility of the other currencies under research.

Further, specifically in the case of foreign exchange another interpretation of asymmetries in volatility spillovers presents itself. The six currencies under research are base currencies. A negative change of the base currency's unit price in terms of the U.S. dollar means that the amount of dollars needed to buy one unit of the base currency is smaller. Thus, a negative change (or negative return) indicates a depreciation of the base currency with respect to the dollar. Spillovers from volatility due to negative returns (and computed with the help of negative realized semivariance $RS^-$) then mean spillovers that emerge due to temporary depreciations of the base currency. A similar logic applies to show that positive realized semivariance ($RS^+$) captures volatility that is due to the positive returns of the base currency, meaning the temporary appreciation of the base currency. We have to stress two issues, though. One, the depreciation or appreciation of a currency is usually understood as a somewhat longer process. Since we employ intra-day data, temporary depreciations and appreciations (negative and positive returns) occur frequently and often move in opposite directions. Hence, they do not represent a longer process from a macroeconomic perspective. Two, it follows that temporary depreciations and appreciations (employed in the form of returns to quantify volatility spillovers) do not necessarily correlate with periods of appreciation or depreciation of a specific currency. Despite the fact that sometimes both events occur simultaneously, it is not a rule. Finally, the illustration of temporary depreciation and appreciation movements behind volatility spillovers is useful for the economic interpretation of our results as well as for comparing our results with related evidence, albeit limited, in the literature. However, by acknowledging its limitations, henceforth we rather employ the standard terminology described and used in the literature; i.e. bad and good volatility spillovers.

Based on the results presented in this subsection we conclude that the net bad volatility spillovers (the SAM in Figure \ref{Fig4}) dominate the good volatility spillovers. Thus, during much of our sample period negative shocks were driving volatility spillovers. Further, there is a difference in the nature of the underlying key factors related to asymmetries in volatility spillovers. Good volatility spillovers of the six currencies under research are linked with (i) the global financial crisis and its subprime crisis nexus in the U.S. and (ii) different monetary policies among key world central banks as well as developments on commodities markets. On the other hand, bad volatility spillovers are chiefly tied to the dragging sovereign debt crisis in Europe. It seems that a combination of monetary and real economy events is behind the net positive spillovers, while fiscal factors are linked with net negative spillovers. Hence, not only the origin of major factors but also their nature can be found behind the asymmetries in volatility spillovers on forex markets.

\subsection{Asymmetries in directional volatility spillovers}
Following the above outline we now proceed with an assessment of asymmetries in directional spillovers among individual currencies. The condensed information on how the asymmetries in directional spillovers propagate is presented in Table \ref{tab:spills2}. The convenient matrix format allows to distinguish proportions in which good and bad volatilities from individual currencies propagate across the market and result in positive and negative spillovers that materialize in the volatilities of the currencies under research. Volatility spillovers that are above average levels might be detected for interactions between commodity currencies (CAD, AUD) and the euro and Swiss franc pair. These patterns also resonate with the non-asymmetric spillovers reported in subsection \ref{sec:ds}. Unfortunately, the condensed table does not reveal the dynamics in the pattern of directional asymmetries. Hence, the full dynamics is presented in graphical form below.

The detailed dynamics is provided in Figure \ref{Fig5}, where we present directional asymmetries in volatility spillovers coming from a specific currency to the rest of the currencies under research. First we show how the bad volatility of a specific currency influences the volatility of the other five currencies in the system (first row). The graphs are calculated from a 12-variable system of six $RS^+$ and six $RS^-$ as a sum of the column in a matrix shown in Table \ref{tab:spills2} excluding all diagonals of all four 6x6 block-matrices in the system. The least pronounced positive spillovers are visible in the case of CAD, which means that relatively small spillovers that are due to positive shocks are transmitted from CAD to other currencies. The remaining evidence points to comparable amounts of positive spillovers coming from the currencies.

In a similar fashion we are able to isolate the effects of bad volatility. In the second row of Figure \ref{Fig5} we plot bad volatility spillovers coming from a specific currency to the rest of the currencies. Most of the negative spillovers come from AUD, CAD, and EUR as their plots reach relatively high levels over the entire time span. On the other hand, the smallest proportion of spillovers due to negative shocks is transmitted from JPY to the other currencies. The rest of the currencies record a comparable extent of negative spillovers transmitted from them. Based on the evidence in the first two rows of Figure \ref{Fig5}, we are able to reject Hypothesis 2 because both negative and positive directional spillovers from each currency are transmitted to the rest of the currencies in the portfolio and these transmissions are not symmetric.

Finally, in the third row, we present $net$ asymmetric directional spillovers constructed as a difference between the values plotted in the first and second rows. Formally, the net asymmetric directional spillovers are defined as the difference of the sums of the columns of $RS^+$ and $RS^-$ in \ref{tab:spills2} excluding all diagonals of all four 6x6 block-matrices in the system. The net asymmetric directional spillovers provide the key interpretation value because they measure whether the good volatility of a specific currency affects the volatility of the other currencies more than bad volatility (positive domain of the plot) or whether net negative spillovers exhibit a greater impact (negative domain of the plot). In sum, the evidence in the third row of Figure \ref{Fig5} points to the fact that volatility spillovers transmitted from one currency exhibit an asymmetric impact on the volatility of the other currencies in portfolio. Thus, we are able to reject Hypothesis 3. We can further gauge that negative spillovers occur more often and with somewhat larger size than positive spillovers. Hence, negative spillovers transmitted from one currency impact the volatility of the other currencies in the portfolio more than positive spillovers. Specific impacts are described below.

Both commodity currencies, AUD and CAD, transmit heavily net negative spillovers to other currencies, especially during 2010 -- 2011 and also well into 2012. Further, while AUD occasionally also transmits net positive spillovers, CAD is by and large on the negative-shocks side and its net transmitting position does not experience any regime break associated with either GFC or the European debt crisis. Large asymmetries in 2011 -- 2012 reflect further increases of prices in commodities markets until they burst in 2012. Decreasing asymmetries for both AUD and CAD around 2013 pair well with developments on commodities markets and with the fact that for that particular period commodities seem to have decoupled from their strong negative correlation with the U.S. dollar.

Vehicle currencies (EUR, JPY) exhibit highly polarized behavior. The exact timings of the worst episodes during the European sovereign debt crisis contour sharply the periods when the EUR transmits net negative spillovers. The U.S.-bred GFC on the other hand coincides with the EUR net positive spillovers. Similarly, the period when the ECB began to buy bonds (2014 -- 2015) is characterized by net positive spillovers as well. The shift in the regime change is quite clear and these key events are most likely behind such asymmetries. The JPY exhibits a different dynamics: diffusion of the net directional spillovers due to positive returns dominates most of the time span. Conversely, the period 2008 -- 2012 exhibits an almost unbroken pattern of net positive spillovers. The customary forex interventions of the Bank of Japan against the currency’s strength are a likely driver of the shocks behind the net spillovers. The pattern (including the interventions) also correlates with the fact that during 2006 -- 2010 many Japanese insurance companies and pension funds engaged in purchases of foreign bonds that further increased pressure on the yen’s value. In addition, the emergence of many small forex brokers also potentially contributed to volatility on the market. A specific event that breaks the pattern can be detected in the asymmetries plot, though. There is a decrease in positive spillovers and even a small swelling of net negative spillovers from JPY to other currencies in the second quarter of 2011. This evidences the effect, albeit marginal, of the joint intervention of the Fed, ECB, Bank of England, and Bank of Canada to assist the Bank of Japan in its effort to defend the yen and harbor its volatility in the aftermath of the devastating earthquake.\footnote{The earthquake off the Pacific coast of the Tōhoku region and the subsequent tsunami occurred on March 11, 2011. It was the most powerful earthquake ever recorded to have hit Japan. Massive damages included the meltdown of three reactors in the Fukushima Daiichi Nuclear Power Plant.} The rest of the researched period from 2012 on is characterized by either net negative spillovers or marginal alternating asymmetries.

Based on the extent of the net spillovers, the non-Eurozone currencies (GBP and CHF) seem to be modest transmitters of net directional spillovers onto other currencies. However, their net spillover plots do not bear much resemblance. Both currencies display qualitative differences in unbroken portions of their net spillovers: net negative spillovers dominate the European debt crisis period for the GBP while in the case of CHF net positive spillovers prevail during the GFC. The Swiss National Bank began to be quite active in 2009 with the aim to weaken its currency. Over 2009 -- 2011 its steps involved forex interventions, verbal interventions, and interest rate adjustments. It is interesting that the lowest net spillovers are visible in 2011, when the bank gave up on limiting the CHF 1.20-per-euro bound and discontinued its managed float policy of \textit{capping}.

The above results on the asymmetries in volatility spillovers are unique in that they represent qualitatively new information. We stated earlier that the literature lacks a proper treatment of asymmetries in volatility spillovers in forex markets. As a result, the single study with which we can compare our results is that of \cite{galagedera2012effect}, who show that during the period of the subprime crisis (2008 -- 2009), the appreciation of the yen against the U.S. dollar had a greater impact on the U.S. dollar-yen volatility spillover than the yen’s depreciation. Our results from the same period fully support their finding (see the third row in Figure \ref{Fig5}). In addition, even later, until early 2012, the pattern does not change as the yen’s positive spillovers, i.e., volatility spillovers computed based on positive returns or temporary appreciation changes, exhibited a larger impact than negative spillovers. The pattern changes only from 2012 on when negative spillovers begin to prevail. Their extent is visibly smaller than that of the positive spillovers, though. \cite{galagedera2012effect} also show that the appreciation and depreciation of the U.S. dollar against the euro does not appear to have an asymmetric effect on the Euro-U.S. dollar volatility spillover. In this case we are cautious with their finding because our results show an asymmetric effect of euro volatility spillovers being transmitted to other currencies. During the investigated subprime crisis period, positive spillovers from the euro (i.e. spillovers due to temporary appreciations) dominate volatility spillovers going from the euro to other currencies.\footnote{We have to stress that, because the methodologies employed in \cite{galagedera2012effect} and in our analysis are different, both sets of results are not directly comparable.}

\section{Conclusion \label{sec:conclusion}}
We extend the procedure of \cite{barunik2016asymmetric} to quantify volatility spillovers that are due to bad and good volatility (proxied by negative and positive returns) to better fit the assessment of volatility spillovers on forex markets. The procedure is based on a computation of the volatility spillover index \citep{diebold2012better} by considering separately negative and positive changes in returns via realized semivariances \citep{shephard2010measuring}. The approach allows us to quantify (total and directional) volatility spillover indices robust to ordering in VAR and to capture asymmetries in volatility spillovers. Due to the non-existing established methodology, which would provide the detailed evidence on the dynamics of the asymmetries in volatility spillovers, our approach brings insights that could not be obtained earlier.

Using high-frequency intra-day data over 2007 -- 2015 we apply the method on a set of the most actively traded currencies quoted against the U.S. dollar, including the Australian Dollar (AUD), British Pound (GBP), Canadian Dollar (CAD), Euro (EUR), Japanese Yen (JPY), and Swiss Franc (CHF). Based on the analysis of these currencies we provide a wealth of detailed results.

We show that the extent of spillover transmission among currencies is not uniform. Each currency's net position, in terms of volatility spillovers it receives or transmits, is quite different: while GBP and CAD are mostly spillover givers, AUD, JPY, and EUR are mostly spillover receivers, and CHF is a balanced currency. Our findings also directly support the presence of heat-wave volatility clustering \citep{engle1990meteor} as there are substantial directional spillovers among currencies within a specific market.

Further, we decisively show that volatility spillovers in the portfolio of currencies exhibit distinctive asymmetries. Such asymmetries are not uniform with respect to currencies, timing, or potential underlying factors. In this respect the negative spillovers dominate positive spillovers in their magnitude as well as frequency; this behavior distinguishes the forex market from stocks and commodities markets where the divide between negative and positive asymmetries is much less prominent \citep{barunik2016asymmetric,barunik2015}. Negative spillovers are chiefly tied to the dragging sovereign debt crisis in Europe. Positive spillovers correlate with the subprime crisis in the U.S. and different monetary policies among key world central banks along with developments on commodities markets. Hence, a combination of monetary and real economy events is behind the net positive asymmetries in volatility spillovers while fiscal factors are linked with net negative spillovers.

Finally, we provide evidence that asymmetries exist also in directional spillovers. We show that currencies do not display a similar pattern in how their net asymmetric directional spillovers propagate -- i.e., the forex market exhibits asymmetric volatility connectedness. It is true that some currencies display a common pattern over a certain subset of the time span, chiefly in connection with major economic or financial events. However, the pattern is not decisively comparable over the entire time span. For example, commodity currencies (CAD, AUD) display a similar pattern with the euro during the major phases of the European sovereign debt crisis. However, all three currencies (CAD, AUD, EUR) transmit net asymmetric spillovers in a remarkably different fashion during the GFC period. In any event, negative directional spillovers transmitted from one currency impact the volatility of other currencies in the portfolio more than positive spillovers. Thus, asymmetric volatility connectedness on the forex market is dominated by negative changes and this sharply differentiates it from, for example, the U.S. stock market.

\newpage
\section*{References}
\bibliographystyle{chicago}
\bibliography{spilloversbib}

\appendix
\section{Realized variance and semivariance \label{sec:semi}}
In this Section we briefly introduce realized measures that we use for volatility spillover estimation. We begin with realized variance and then we describe realized semivariances. Realized measures are defined on a continuous-time stochastic process of log-prices, $p_t$, evolving over a time horizon $[0\le t \le T]$. The process consists of a continuous component and a pure jump component,
\begin{equation}
p_t=\int_0^t\mu_s ds + \int_0^t\sigma_s d W_s + J_t,
\end{equation}
where $\mu$ denotes a locally bounded predictable drift process, $\sigma$ is a strictly positive volatility process, and $J_t$ is a jump part, and all is adapted to some common filtration $\mathcal{F}$. The quadratic variation of the log prices $p_t$ is:
\begin{equation}
\label{eq:qv}
[p_t,p_t] = \int_0^t\sigma_s^2 ds+\sum_{0<s\le t}(\Delta p_s)^2,
\end{equation}
where $\Delta p_s = p_s - p_{s-}$ are jumps, if present. The first component of Eq. (\ref{eq:qv}) is integrated variance, whereas the second term denotes jump variation. \cite{andersen1998answering} proposed estimating quadratic variation as the sum of squared returns and coined the name ``realized variance" ($RV$). The estimator is consistent under the assumption of zero noise contamination in the price process.

Let us denote the intraday returns $r_k=p_k-p_{k-1}$, defined as a difference between intraday equally spaced log prices $p_0,\ldots,p_n$ over the interval $[0,t]$, then
\begin{equation}
RV=\sum_{k=1}^n r_k^2
\end{equation}
converges in probability to $[p_t,p_t]$ with $n\rightarrow \infty$.

Lately, \cite{shephard2010measuring} decomposed the realized variance into realized semivariances ($RS$) that capture the variation due to negative ($RS^-$) or positive ($RS^+$) price movements (e.g., bad and good volatility). The realized semivariances are defined as:
\begin{eqnarray}
RS^-&=& \sum_{k=1}^n \mathbbm{I}(r_k<0) r_k^2, \\
RS^+&=& \sum_{k=1}^n \mathbbm{I}(r_k\ge0) r_k^2.
\end{eqnarray}
Realized semivariance provides a complete decomposition of the realized variance, hence:
 \begin{equation}
RV=RS^- + RS^+.
\end{equation}
The limiting behavior of realized semivariance converges to $1/2\int_0^t\sigma_s^2 ds$ plus the sum of the jumps due to negative and positive returns \citep{shephard2010measuring}. The negative and positive semivariance can serve as a measure of downside and upside risk as it provides information about variation associated with movements in the tails of the underlying variable.

\section{Tables}

\begin{table}[h]
\scriptsize
\caption{Spillover matrix ($2N \times 2N$)}
\centering
\begin{tabular}{rrcccccccccccc}
&&\multicolumn{6}{c}{\textbf{$RS^+$}} &\multicolumn{6}{c}{\textbf{$RS^-$}}\\
&& AUD & GBP & CAD & EUR & JPY & CHF & AUD & GBP & CAD & EUR & JPY & CHF \\
{\multirow{6}{*}{\rotatebox[origin=c]{90}{\textbf{$RS^+$}}}}
&AUD & $\mathbf{\omega_{1,1}}$ & $\omega_{1,2}$ & $\omega_{1,3}$ & $\omega_{1,4}$ & $\omega_{1,5}$ & $\omega_{1,6}$ & $\mathbf{\omega_{1,7}}$ & $\omega_{1,8}$ & $\omega_{1,9}$ & $\omega_{1,10}$ & $\omega_{1,11}$ & $\omega_{1,12}$ \\
&GBP & $\omega_{2,1}$ & $\mathbf{\omega_{2,2}}$ &.&.&.&.&.&$\mathbf{\omega_{2,8}}$&.&.&.&.\\
&CAD &$\omega_{3,1}$&.&$\mathbf{\omega_{3,3}}$&.&.&.&.&.&$\mathbf{\omega_{3,9}}$&.&.&.\\
&EUR &$\omega_{4,1}$&.&.&$\mathbf{\omega_{4,4}}$&.&.&.&.&.&$\mathbf{\omega_{4,10}}$&.&.\\
&JPY &$\omega_{5,1}$&.&.&.&$\mathbf{\omega_{5,5}}$&.&.&.&.&.&$\mathbf{\omega_{5,11}}$&.\\
&CHF &$\omega_{6,1}$&.&.&.&.&$\mathbf{\omega_{6,6}}$&.&.&.&.&.&$\mathbf{\omega_{6,12}}$\\
\\
{\multirow{6}{*}{\rotatebox[origin=c]{90}{\textbf{$RS^-$}}}}
&AUD &$\mathbf{\omega_{7,1}}$&.&.&.&.&.&$\mathbf{\omega_{7,7}}$&$.$&.&.&.&.\\
&GBP & $\omega_{8,1}$ & $\mathbf{\omega_{8,2}}$ &.&.&.&.&$.$&$\mathbf{\omega_{8,8}}$&.&.&.&.\\
&CAD &$\omega_{9,1}$&.&$\mathbf{\omega_{9,3}}$&.&.&.&.&.&$\mathbf{\omega_{9,9}}$&.&.&.\\
&EUR &$\omega_{10,1}$&.&.&$\mathbf{\omega_{10,4}}$&.&.&.&.&.&$\mathbf{\omega_{10,10}}$&.&.\\
&JPY &$\omega_{11,1}$&.&.&.&$\mathbf{\omega_{11,5}}$&.&.&.&.&.&$\mathbf{\omega_{11,11}}$&.\\
&CHF &$\omega_{12,1}$&.&.&.&.&$\mathbf{\omega_{12,6}}$&.&.&.&.&.&$\mathbf{\omega_{12,12}}$\\
\\
\end{tabular}
\label{tab:spillsetting}
\end{table}

\begin{table}[h]
\small
\caption{Volatility spillover table for $N$-dimensional VAR model.}
\centering
\begin{tabular}{lrrrrrrr}
\toprule
& AUD & GBP & CAD & EUR & JPY & CHF& \textbf{FROM}\\
\cmidrule{2-7}
AUD &\textbf{31.39}&16.42&16.00&13.58&12.05&10.55&68.61\\
GBP &15.96&\textbf{29.51}&13.72&17.10&11.19&12.52&70.49\\
CAD &19.43&16.91&\textbf{30.03}&12.51&10.24&10.89&69.97\\
EUR &14.57&17.62&10.59&\textbf{28.73}&8.76&19.73&71.27\\
JPY &13.34&14.96&9.17&10.89&\textbf{40.15}&11.50&59.85\\
CHF&12.85&14.47&10.41&22.01&10.44&\textbf{29.83}&70.17\\
\cmidrule{2-7}
\textbf{TO} &76.15&80.37&59.88&76.09&52.68&65.19&\textbf{TOTAL}\\
        & &   &&&  &   & \textbf{68.39} \\
\bottomrule
\end{tabular}
\label{tab:spills}
\end{table}

\begin{table}
\scriptsize
\caption{Descriptive statistics for $\sqrt{RS^+}$ and $\sqrt{RS^-}$}
\centering
\begin{tabular}{lrrrrrrrrrrrrr}
\toprule
&\multicolumn{6}{c}{\textbf{$\sqrt{RS^+}$}} & \multicolumn{6}{c}{\textbf{$\sqrt{RS^-}$}}\\
 & AUD & GBP & CAD & EUR & JPY & CHF & AUD & GBP & CAD & EUR & JPY & CHF \\
 \cmidrule{2-7}
 \cmidrule{8-13}
 mean&0.0059&0.0041&0.0046&0.0043&0.0045&0.0048&0.006&0.0042&0.0045&0.0043&0.0045&0.0048\\
stdev &0.0035&0.0022&0.0022&0.0019&0.0024&0.0022&0.0037&0.0022&0.0022&0.0019&0.0023&0.0021\\
skew &3.3795&2.427&1.7602&1.7531&2.6455&2.6323&3.5585&2.5774&1.7591&1.5714&2.7584&2.3364\\
kurt &21.0783&11.5377&8.2714&8.634&16.4973&19.3321&24.1825&13.4676&8.0295&7.1948&19.1172&15.3637\\
min &0.0016&0.0011&0.001&0.0009&0.0011&0.001&0.0015&0.0012&0.0013&0.0011&0.001&0.0011\\
max &0.0394&0.0198&0.0205&0.0178&0.0299&0.0297&0.0475&0.0249&0.0189&0.0173&0.0314&0.0272\\
\bottomrule
\end{tabular}
\label{tab:descr}
\end{table}

\begin{table}
\scriptsize
\caption{Volatility spillover table for the $2N$-dimensional VAR model with realized semivariances}
\centering
\begin{tabular}{lrrrrrrrrrrrrrrr}
\toprule
&&\multicolumn{6}{c}{\textbf{$RS^+$}} &&\multicolumn{6}{c}{\textbf{$RS^-$}}\\
&& AUD & GBP & CAD & EUR & JPY & CHF && AUD & GBP & CAD & EUR & JPY & CHF & \textbf{FROM}\\
\cmidrule{3-8}\cmidrule{10-15}
{\multirow{6}{*}{\rotatebox[origin=c]{90}{\textbf{$RS^+$}}}}
&AUD &\textbf{15.97}&7.55&7.22&7.04&7.28&6.04 && \textbf{14.62}&8.85&9.30&6.40&4.88&4.86&69.41\\
&GBP &8.15&\textbf{15.16}&6.17&8.99&6.69&7.29 && 7.70&\textbf{13.92}&7.54&7.85&4.84&5.71&70.92\\
&CAD &9.59&7.47&\textbf{16.39}&6.14&5.85&5.69 && 9.66&9.14&\textbf{14.16}&6.18&4.49&5.26&69.45\\
&EUR &7.79&8.53&4.95&\textbf{16.52}&5.68&12.27 && 6.61&8.64&5.65&\textbf{11.65}&3.63&8.08&71.83\\
&JPY &6.42&7.33&3.94&6.36&\textbf{26.06}&7.45 && 6.85&6.97&5.27&4.00&\textbf{15.32}&4.03&58.62\\
&CHF &7.07&7.56&4.89&13.15&6.85&\textbf{18.77} && 5.76&6.65&5.46&8.53&4.14&\textbf{11.17}&70.06\\
\\
{\multirow{6}{*}{\rotatebox[origin=c]{90}{\textbf{$RS^-$}}}}
&AUD &\textbf{11.59}&6.22&5.77&5.54&7.01&4.85 && \textbf{19.98}&9.68&10.32&7.86&5.52&5.66&68.42\\
&GBP &7.07&\textbf{9.99}&5.48&7.02&6.12&5.60 && 8.96&\textbf{18.93}&8.72&9.84&5.31&6.96&71.08\\
&CAD &8.63&6.61&\textbf{10.67}&5.30&5.86&5.20 && 11.00&9.88&\textbf{18.81}&7.41&4.67&5.98&70.53\\
&EUR &6.44&6.57&4.18&\textbf{10.67}&4.26&8.00 && 8.36&10.89&7.02&\textbf{17.86}&4.24&11.51&71.47\\
&JPY &5.88&6.86&3.78&5.39&\textbf{19.65}&6.06 && 7.38&7.75&5.54&5.10&\textbf{21.25}&5.38&59.10\\
&CHF &6.09&5.95&4.47&9.19&5.06&\textbf{12.75} && 6.90&8.32&6.51&12.45&5.17&\textbf{17.14}&70.11\\
\cmidrule{3-8}\cmidrule{10-15}
&\textbf{TO} & 73.11&70.64&50.85&74.13&60.63&68.45 && 79.19&86.76&71.32&75.62&46.87&63.42&\textbf{TOTAL}\\
&&&&&&&&&&&&&&&\textbf{68.42}\\
\bottomrule
\end{tabular}
\label{tab:spills2}
\end{table}

 \pagebreak
 \newpage
 \addtolength{\topmargin}{-0.5in}
 \setlength{\oddsidemargin}{0in}
 \setlength{\evensidemargin}{0in}
 \setlength{\textheight}{9.4in}

 \section{Figures}

\begin{landscape}
  \begin{figure}[!ht]
  \center
    \includegraphics[scale= 0.55]{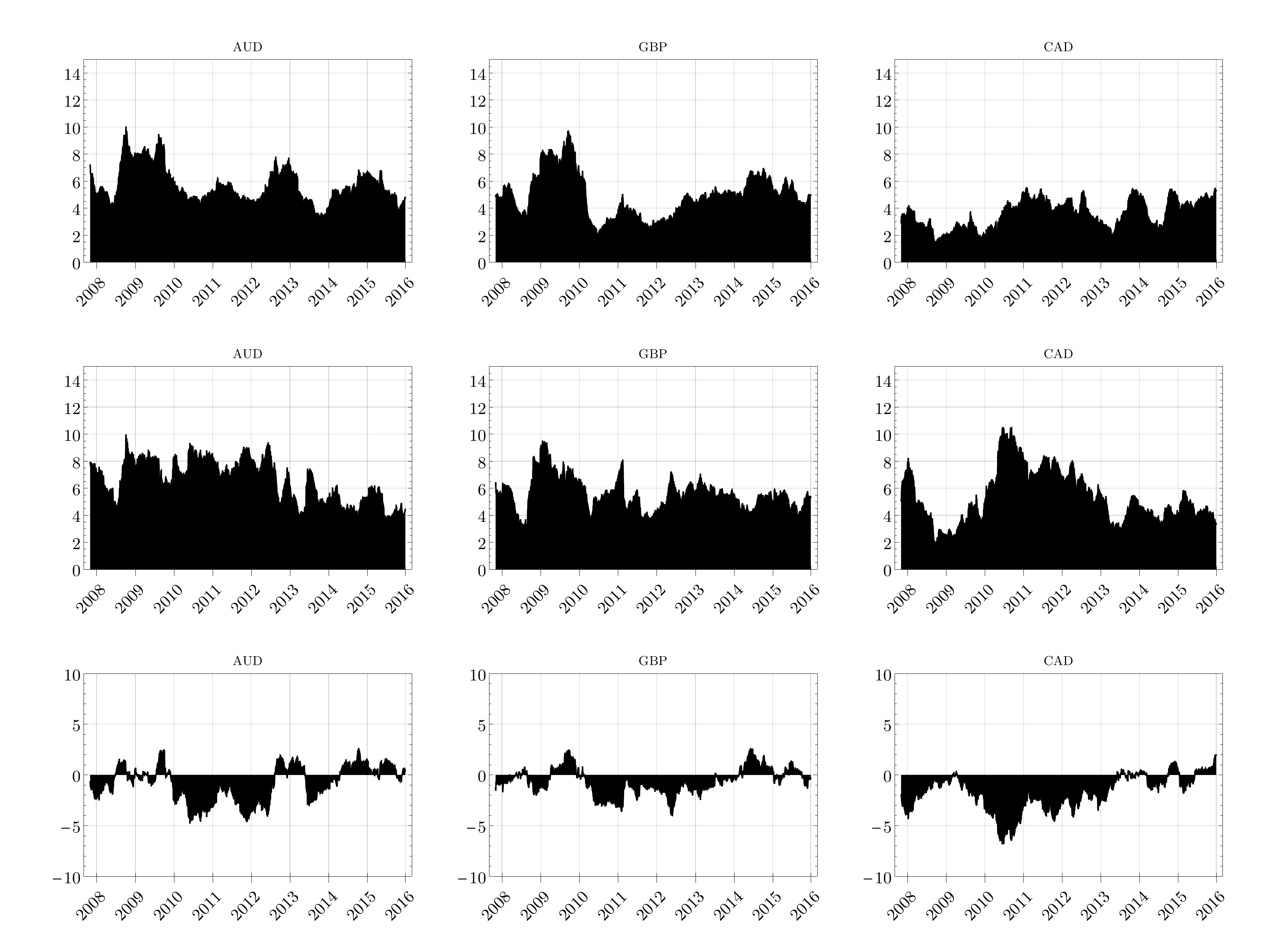}
    \caption{upper panel: directional spillovers, direction to, from good volatility, middle panel: directional spillovers, direction to, from bad volatility, lower panel: the directional spillover asymmetry measure $\mathcal{SAM}^H_{2N,i\rightarrow\bullet}$}
    \label{Fig5}
  \end{figure}

  \begin{figure}[!ht]
    \ContinuedFloat
    \center
  \includegraphics[scale= 0.55]{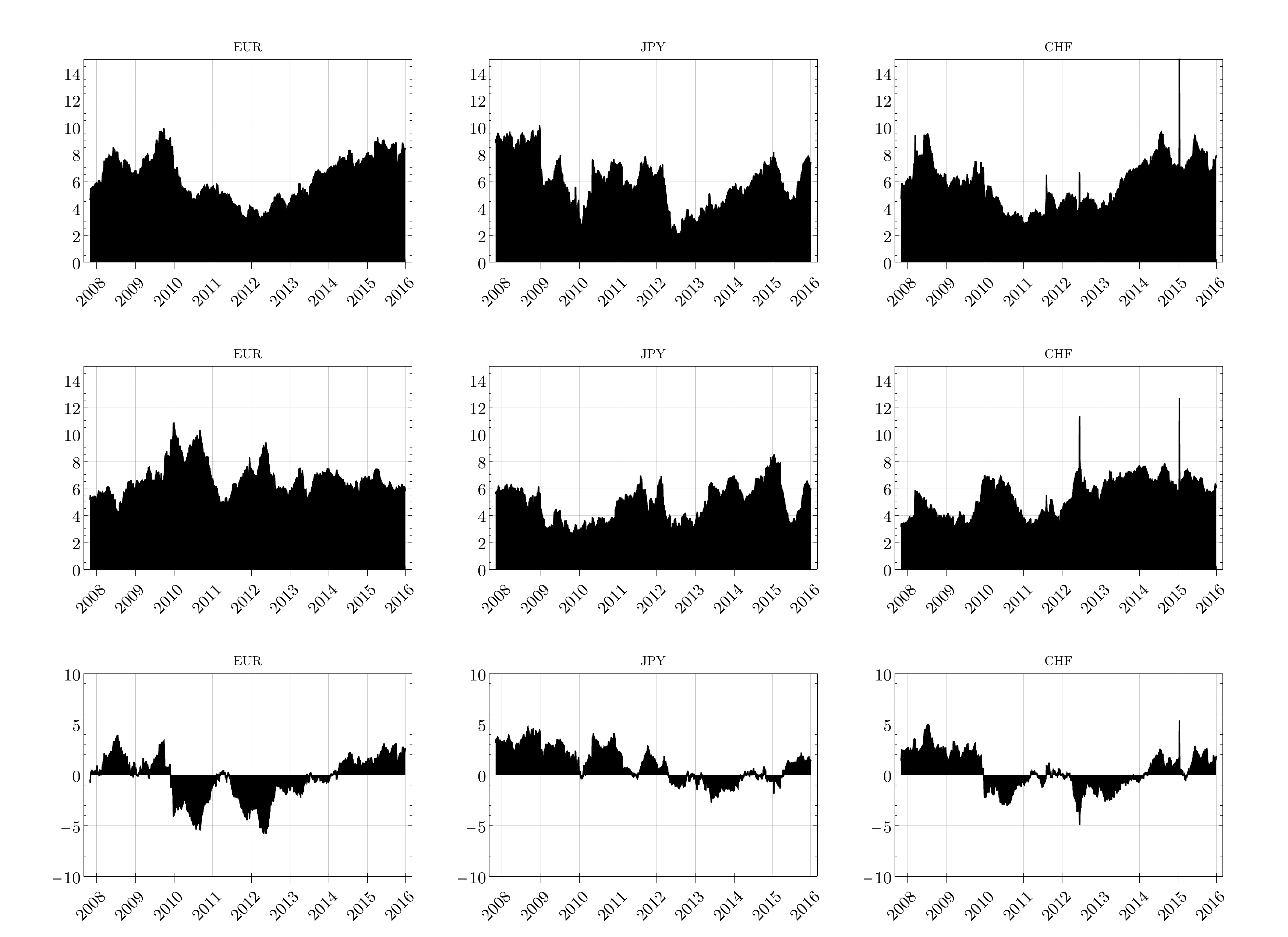}
    \caption{upper panel: directional spillovers, direction to, from good volatility, middle panel: directional spillovers, direction to, from bad volatility, lower panel: the directional spillover asymmetry measure $\mathcal{SAM}^H_{2N,i\rightarrow\bullet}$ (cont.)}
    \end{figure}
\end{landscape}

\end{document}